\documentclass[prl,twocolumn,showpacs,amsmath,amssymb,superscriptaddress]{revtex4}
\usepackage{graphicx}
\usepackage{dcolumn}
\usepackage{bm}
\usepackage{epsfig}

\begin{document}


\title{Magnetism in Geometrically
Frustrated YMnO$_{3}$ Under Hydrostatic Pressure Studied with Implanted Muons}

\author{T. Lancaster}
\email{t.lancaster1@physics.ox.ac.uk}
\author{S.J. Blundell}
\affiliation{
Clarendon Laboratory, Oxford University Department of Physics, Parks
Road, Oxford, OX1 3PU, UK
}

\author{D. Andreica}
\affiliation{
Laboratory for Muon Spin Spectroscopy, Paul Scherrer Institut, CH-5232
Villigen, Switzerland}
\altaffiliation{On leave from Faculty of Physics, Babes-Bolyai University,
400084 Cluj-Napoca,  Romania}

\author{M. Janoschek}
\affiliation{Technische Universit{\"a}t M{\"u}nchen, Physics
  Department E21, D-85747 Garching, Germany}
\affiliation{
Laboratory for Neutron Scattering, ETHZ and Paul Scherrer Institut,
CH-5232 Villigen, Switzerland}

\author{B. Roessli} 
\author{S.N. Gvasaliya}
\affiliation{
Laboratory for Neutron Scattering, ETHZ and Paul Scherrer Institut,
CH-5232 Villigen, Switzerland}

\author{K. Conder}
\affiliation{
Laboratory for Developments and Methods, Paul Scherrer Institut, CH-5232
Villigen, Switzerland}

\author{E. Pomjakushina}
\affiliation{
Laboratory for Neutron Scattering, ETHZ and Paul Scherrer Institut,
CH-5232 Villigen, Switzerland}
\affiliation{
Laboratory for Developments and Methods, Paul Scherrer Institut, CH-5232
Villigen, Switzerland}

\author{M.L. Brooks}
\author{P.J. Baker}
\author{D. Prabhakaran}
\author{W. Hayes}
\affiliation{
Clarendon Laboratory, Oxford University Department of Physics, Parks
Road, Oxford, OX1 3PU, UK
}
\author{F.L. Pratt}
\affiliation{
ISIS Facility, Rutherford Appleton Laboratory, Chilton, Oxfordshire OX11 0QX, UK}

\date{\today}

\begin{abstract}
The ferroelectromagnet YMnO$_{3}$ consists
of weakly coupled triangular layers of $S=2$ spins. 
Below $T_{\mathrm{N}}\approx 70$~K muon-spin relaxation data show
two oscillatory relaxing
signals due to magnetic order, with no purely relaxing signals 
resolvable (which would require different coexisting spin
distributions).
The transition temperature 
$T_{\mathrm{N}}$ increases with applied hydrostatic pressure,
even though the ordered moment decreases. These results suggest
that pressure 
increases both the exchange coupling between the layers and the
frustration within the layers.
\end{abstract}

\maketitle

The hexagonal $R$MnO$_{3}$ manganites \cite{yakel}
($R$=Ho, Er, Tm, Yb, Lu, Y or Sc) 
are a class of magnetically ordered materials that 
also possess ferroelectric properties and some degree of
 magnetoelectric coupling, 
suggesting that an understanding of the magnetism may allow
manipulation of the electric polarization with possible
device applications \cite{eerenstein}.
Moreover, these compounds are layered and exhibit geometric 
frustration within their layers, offering the possibility
of studying the effect of competing interactions in low-dimensional
systems \cite{moessner}.

YMnO$_{3}$, an insulator which undergoes a ferroelectric transition
at 
$T_{\mathrm{E}}= 913$~K \cite{frohlich},
is the most intensively studied of the $R$MnO$_{3}$ series.
The magnetic system is based on a frustrated architecture, 
with  Mn$^{3+}$ ($S=2$) ions forming a two-dimensional (2D) 
 corner-sharing triangular network. 
Studies of magnetic
susceptibility \cite{katsufuji}
confirm the frustrated nature of the system, with a large ratio of
Weiss temperature ($|\Theta|=705$~K) 
to antiferromagnetic (AFM) ordering temperature ($T_{\mathrm{N}} \approx 70$~K).
Neutron diffraction
studies \cite{munoz,brown} have found that below 
 $T_{\mathrm{N}}$, the Mn$^{3+}$ spins lie in the $ab$ plane
and adopt a 120$^{\circ}$ structure, with a Mn moment of 
2.9--3.1$\mu_{\mathrm{B}}$ at 1.7~K.
This is below the expected value of 4$\mu_{\mathrm{B}}$, 
due to fluctuations associated with either the frustration or the low
dimensionality. 
Heat capacity measurements \cite{tomuta,katsufuji} suggested
incomplete ordering of the Mn spins below $T_{\mathrm{N}}$ 
and elastic and inelastic neutron scattering (INS)
measurements \cite{park} found strong diffuse scattering persisting
across $T_{\mathrm{N}}$; this was taken as evidence for 
a spin liquid phase which coexists with the ordered phase below
$T_{\mathrm{N}}$. However, it has been claimed that by taking low
energy Einstein modes into account, heat capacity data may be
consistent with conventional AFM ordering \cite{tachibana}.
Other INS studies \cite{sato,roessli} have found evidence for 
coexisting three-dimensional (3D) and 2D fluctuations.
More recent neutron studies \cite{janoschek,kozlenko} 
 show
that the ordered moment decreases with increasing hydrostatic
pressure. This observation was explained in terms of a
pressure-induced
change in volume
fraction of ordered and spin liquid components of the material \cite{kozlenko}.

In this paper we present the results of muon-spin relaxation \cite{steve} 
($\mu^{+}$SR) measurements made on YMnO$_{3}$
 at ambient pressure and as a function of hydrostatic pressure 
$p$ up to $p=13.7$~kbar. Muons are a sensitive local probe of the 
spin distributions in a magnetic material and 
have proven particularly useful in
probing frustration related effects \cite{frust};
however, measurements made under pressure have been less common. 
Strikingly, we show that $T_{\mathrm{N}}$ 
increases with increasing $p$, even though the magnetic moment decreases.
This provides, as we shall show,
 an insight into the role of the finely balanced
interactions in this system.

\begin{figure}
\begin{center}
\epsfig{file=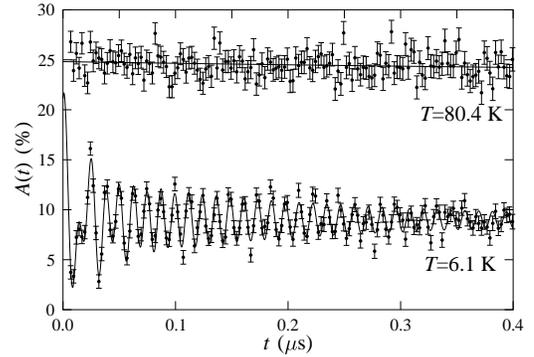,width=6.9cm}
\caption{ZF $\mu^{+}$SR spectra measured at $T=6.1$~K
and 80.4~K at $p=1$~bar, with oscillations clearly
observable below $T_{\mathrm{N}}$. 
\label{data}}
\end{center}
\end{figure}

Zero-field (ZF) $\mu^{+}$SR measurements \cite{time}
were made on a single crystal
sample of YMnO$_{3}$ using the GPS instrument
at the Swiss Muon Source (S$\mu$S) and on the MuSR instrument at the 
ISIS facility. The muon 
spin was directed 
along the crystallographic $c$-axis of the sample.
Transverse field (TF) $\mu^{+}$SR measurements were made 
on a powder sample of YMnO$_{3}$
under hydrostatic pressure using 
high energy incident muons on the $\mu$E1 decay muon beam line
at S$\mu$S.
For these measurements, the sample was packed into a cylinder
approximately 
7~mm in diameter and 
18~mm in length which was mounted in a Cu-Be25 piston cylinder
pressure-cell with Daphne 
oil used as the pressure medium. 
The pressure was measured by monitoring the 
superconducting transition temperature 
of an
indium wire located in the sample space \cite{samples}.

Example ZF $\mu^{+}$SR spectra at ambient pressure (measured at S$\mu$S)
are shown in Fig.~\ref{data}. Below $T_{\mathrm{N}}$
we observe oscillations in the time dependence of the muon
polarization (the ``asymmetry'' \cite{steve})
characteristic of a quasi-static local magnetic field at the 
muon stopping site. This local field causes a coherent precession of the
spins of those muons with a component of their spin polarization
perpendicular to this local field. 
The frequency of the oscillations is given by
$\nu_{i} = \gamma_{\mu} B_{i}/2 \pi$, where $\gamma_{\mu}$ is the muon
gyromagnetic ratio ($=2 \pi \times 135.5$~MHz T$^{-1}$) and $B_{i}$
is the average magnitude of the local magnetic field at the $i$th muon
site. Any fluctuation in magnitude of these fields will
result in a relaxation of the signal, described by
relaxation rates $\lambda_{i}$. 
Two separate
frequencies were identified in the low temperature spectra,
corresponding to two magnetically inequivalent muon stopping sites in the
material. The larger frequency $\nu_{1}$ is found to have a small
relaxation rate $\lambda_{1}$ while the smaller frequency
$\nu_{2}$ is associated with a relaxation rate $\lambda_{2}$
which is an order of magnitude larger.
The spectra were found to be well fitted using only 
oscillatory components.
 In the ordered phase of YMnO$_{3}$ the Mn$^{3+}$ moments adopt the
 120$^{\circ}$ structure, where the ordered Mn moments lie
within the $a$-$b$ plane \cite{munoz}. 
Although the
initial muon polarization is directed parallel to the $c$-direction,
the fact that the muon couples to dipole fields means that, in
addition to the magnetic field components directed perpendicular
to the muon spin, there may also exist components parallel to the 
muon spin. In our measurements these components only
give rise to a constant background offset. 

To follow the temperature evolution of the observed features
the S$\mu$S spectra  below $T_{\mathrm{N}}$ were
fitted to the functional form
\begin{equation}
\label{fitfunction}
A(t)=A_{\mathrm{bg}} + \sum_{i=1}^{2} A_{i} \mathrm{e}^{\lambda_{i} t} 
\cos (2 \pi \nu_{i} t + \phi_{i}),
\end{equation}
where $A_{\mathrm{bg}}$ represents a constant background contribution,
including the signal from those muons that stop in the silver sample holder or
cryostat tails. Nonzero phases $\phi_{i}$ were required
to fit the observed oscillations because of the
difficulty in resolving features at early times in the spectra
due, at least in part, to the fast initial depolarization feature. 
To give the best fit across the entire
temperature range we fixed $A_{1}= 4.0\%, \phi_{1}=-34^{\circ}$
$A_{2}= 8.8\%$ and $\phi_{2}=-19^{\circ}$.

\begin{figure}
\begin{center}
\epsfig{file=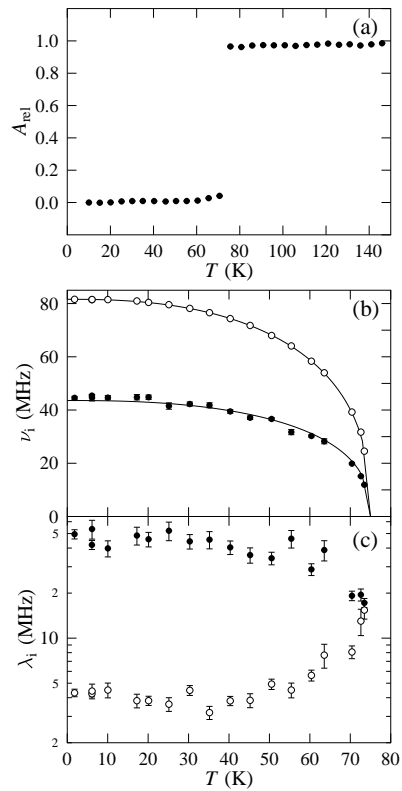,width=6.2cm}
\caption{
(a) Relaxing asymmetry (normalised) $A_{\mathrm{rel}}$ (measured at ISIS)
 showing a step-like change at $T_{\mathrm{N}}$ (see text).
(b) and (c) Fits of the S$\mu$S spectra at $p=1$~bar
to Eq.\ref{fitfunction}. 
(b) Temperature dependence of precession frequencies $\nu_{1}$ (open
circles) and $\nu_{2}$ (closed circles). The solid line is a fit to 
 $\nu_{i}(T)
=\nu_{i}(0)(1-(T/T_{\mathrm{N}})^{\alpha})^{\beta}$
(see main text).
(c) Relaxation rates $\lambda_{1}$ (open
circles) and $\lambda_{2}$ (closed circles). 
\label{fit}}
\end{center}
\end{figure}

We are unable to resolve a relaxing signal due to any
coexisting spin distribution such as the spin liquid phase suggested
to persist into the ordered phase \cite{park}. 
Our muon data are straightforwardly accounted for by a model assuming
conventional long range magnetic order throughout the bulk of the
sample.  The full muon asymmetry is observed above $T_{\rm N}$ and
relaxes exponentially with a single relaxation rate, as expected for a
conventional paramagnetic state.  Below $T_{\rm N}$ we observe an
oscillatory signal with only weak relaxation, as expected for a
well-defined magnetically ordered state.  Thus we find no evidence for
any static magnetic inhomogeneities in our sample, nor any evidence for
coexisting ordered and disordered volume fractions.  
Furthermore, simulations of the internal field distribution expected
from ``droplets'' of spin liquid dispersed in an antiferromagnetic
medium predict a sizeable slowly relaxing fraction, inconsistent
with our data. 
This effectively
rules out a model in which long range order and a disordered spin liquid
somehow coexist in different regions of the sample.  Thus the diffuse
scattering observed in neutron experiments \cite{park} (which has been
proposed to arise from a spin liquid state) results not from phase
separated regions but more likely from high frequency fluctuations which
are motionally narrowed on the muon timescale and which do not affect
the magnetic ground state of the system.  Our data do show a small
missing fraction of relaxing asymmetry (approximately 10\% of the total
signal) corresponding to muons which are depolarized within 1\,ns  of
implantation (not resolvable in our measurements) but this feature only
appears below $T_{\rm N}$, unlike the diffuse scattering which persists
across the transition \cite{park}. 

Complementary data were measured over a longer time window
at the ISIS facility, where the
limited time
resolution does not allow us to resolve oscillations.
Instead, we see
a sharp decrease in the relaxing amplitude $A_{\mathrm{rel}}$ 
as the material 
is cooled through $T_{\mathrm{N}}$.
This is because the local field in the ordered state will 
strongly depolarize the muon-spin if local field 
components are perpendicular 
to the initial muon-spin polarization (or have no effect on it if local field
components are parallel to the initial muon-spin polarization),
removing the relaxing asymmetry from the spectrum.
These data (Fig.~\ref{fit}(a)) show that the relaxing
amplitude does not vary with temperature apart from at
$T_{\mathrm{N}}$,
showing that there is no temperature variation in the volume fractions 
due to the ordered magnetic state (for $T<T_{\mathrm{N}}$) or paramagnetic
state ($T>T_{\mathrm{N}}$).  Moreover, only a weak, temperature
  independent relaxation ($\approx 0.015$~MHz) is observed in
  the ISIS data below $T_{\mathrm{N}}$,
  which is well within the
  ordinarily expected background contribution, confirming
  that no purely relaxing component is needed in Eq.~(\ref{fitfunction}).

Fig~\ref{fit}(b) shows that $\nu_{1}$ and $\nu_{2}$ can be fitted 
by 
$\nu_{i}(T) =\nu_{i}(0)(1-(T/T_{\mathrm{N}})^{\alpha})^{\beta}$
from which
we estimate $T_{\mathrm{N}}=74.7(3)$~K, $\alpha\approx 2.5$
and $\beta = 0.35(3)$,  consistent with 3D Heisenberg 
or 3D $XY$ behaviour, as found in specific heat studies \cite{tachibana}.
Our determination of $\nu_{i}(0)$ allows us to attempt to identify the
muon sites in YMnO$_{3}$. Dipole fields were 
calculated in a sphere containing $\sim 10^5$ Mn ions with
moments of $2.9\mu_{\mathrm{B}}$ arranged in
the $120^{\circ}$ structure.
The positive muon's
position is usually in the vicinity of electronegative O$^{2-}$ ions
\cite{brewer}. 
Candidate muon sites giving rise to the higher oscillation frequency
$\nu_{1}$ are
found to be separated from an O(4) oxygen by 1~\AA\ along the
$c$-direction. This gives sites at coordinates (1/3, 2/3, $z$) and
(2/3, 1/3, $z$), where $z \approx 0.09,\ 0.42,\ 0.59$ and $ 0.92$. 
Several candidate sites for the lower frequency $\nu_{2}$  
are found
close to the planes of triangularly arranged O(1) and O(2)
atoms. One 
possibility is for the sites to lie between oxygens, with the sites
forming
triangles centred again on the (1/3, 2/3, $z$) and (1/3, 2/3, $z$)
positions, where now $z\approx 0.18$ and $\approx 0.32$. 

The division of the muon sites into a set lying close to the
Mn planes and a set between these planes may explain the
difference in the observed relaxation rates. 
The relaxation rates are expected to vary 
as $\lambda \sim \Delta^{2} \tau$, where
$\Delta$ is the second moment of the local magnetic field
distribution and $\tau$ is its fluctuation time.  
Sites giving rise to the frequency $\nu_{1}$
lie close to the Mn planes in well defined positions such that
$\Delta$ is small.  We also expect muons
at these sites to be sensitive to both in-plane (2D) and out-of-plane
(3D) magnetic fluctuations. Relaxation rate  
$\lambda_{1}$ increases as $T_{\mathrm{N}}$ is approached from 
below (Fig.~\ref{fit}(c)) because $\tau$ increases due to the onset 
of critical 
fluctuations close to
the phase transition. 
Sites associated with frequency $\nu_{2}$ lie between the Mn
planes in several positions where there is some variation of the dipole
fields around 40~MHz and consequently a large value of $\Delta$.
 These will be less sensitive
to 2D fluctuations than those lying close to the planes, reducing the
influence of any variation in $\tau$. The 
temperature evolution of $\lambda_{2}$ (Fig.~\ref{fit}(c)) is now
dominated 
by the 
magnitude of $\Delta$, which
scales with the size of the local field. Relaxation rate
$\lambda_{2}$ therefore decreases as the magnetic
transition is approached from below. 

\begin{figure}
\begin{center}
\epsfig{file=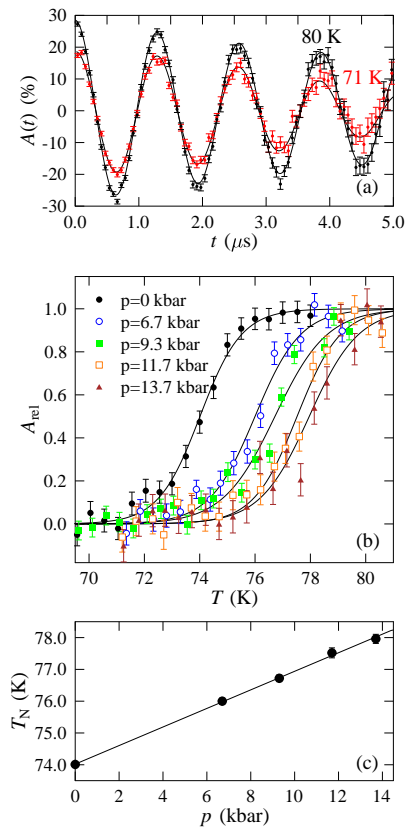,width=5.5cm}
\caption{(Color online.) Results of TF $\mu^{+}$SR measurements made under
hydrostatic pressure. 
(a) Example spectra measured at $p=11.7$~kbar
in a TF of 5~mT at $T=71$~K (red circles) and  $T=80$~K (black circles).
(b) Temperature dependence of the normalised
  amplitude, $A_{\mathrm{rel}}$, for several applied pressures. 
(c) The transition temperature $T_{\mathrm{N}}$ is seen to
increase linearly with increasing pressure. \label{pres}}
\end{center}
\end{figure}

In contrast to the ZF data measured at ambient pressure at S$\mu$S,
it is not possible to resolve precession frequencies in S$\mu$S ZF 
data measured under hydrostatic pressure. This is because the signal 
from the sample amounts to only
20\% of the total measured spectrum from the pressure cell
and is strongly depolarized. To 
observe the magnetic transition,
it was necessary to follow the amplitude of the muon precession 
in a transverse field of $B_{\mathrm{t}}=5$~mT (Fig.~\ref{pres}(a))
 and fit the data to the functional form 
$A(t)=[A_{0} + (A_{\infty}-A_{0})A_{\mathrm{rel}}] e^{-\Lambda t}
\cos (\gamma_{\mu} B_{\mathrm{t}} t)$, where $A_{0}$ and $A_{\infty}$
are found to be independent of pressure, $\Lambda$ is a relaxation
rate and $A_{\mathrm{rel}}$ is the normalized amplitude. The
temperature dependence of $A_{\mathrm{rel}}$ characterizes the
magnetic transition and is shown in Fig.~\ref{pres}(b) for several
applied pressures. The pressure independence of $A_{0}$ and
$A_{\infty}$
are inconsistent with an earlier speculation \cite{kozlenko} that there
is a change in volume fraction between spatially separated 
spin-liquid and AFM states when the pressure is varied. 
To extract the transition temperature we use the phenomenological form
$A_{\mathrm{rel}}= [1+\exp(-b  \{T-T_{\mathrm{N}}\} ) ]^{-1}$
where $b$ is a parameter describing the ``width'' of the transition. 
At all pressures the parameter $b$ remained practically
unchanged at $b \approx 1.2$~K$^{-1}$, while $T_{\mathrm{N}}$ shows
a linear increase with pressure as shown in Fig.~\ref{pres}(c).
A straight line fit yields $T_{\mathrm{N}} = 74.0+0.29 p$, 
where $p$ is the pressure in kbar.
Our value for $\mathrm{d}T_{\mathrm{N}}/\mathrm{d}p = 
0.29$~Kkbar$^{-1}$ is in excess of that 
predicted by the ``10/3'' law \cite{bloch}, 
$\mathrm{d}T_{\mathrm{N}}/\mathrm{d}p = (10/3)T_{\mathrm{N}}/\mathcal{B}$ (where
$\mathcal{B}$ is the bulk modulus), which holds for many oxides and garnets. 
Using $\mathcal{B}=1.65$~Mbar \cite{posadas}
yields an estimate 
$\mathrm{d}T_{\mathrm{N}}/\mathrm{d}p =0.14$~Kkbar$^{-1}$, which is
approximately half the measured value. This strong pressure
dependence of $T_{\mathrm{N}}$ demonstrates the sensitivity of
$T_{\mathrm{N}}$
to 
pressure-induced small changes in 
intralayer $J$ and interlayer coupling $J'$
(see e.g.\ Ref.~\onlinecite{lynn}). 

The increase of $T_{\mathrm{N}}$ with pressure in YMnO$_{3}$
is surprising given that previous studies \cite{kozlenko,janoschek} 
have shown that the magnitude of the ordered moment decreases
with applied pressure, implying an increase in the spin fluctuations
that reduce the value of the magnetic moment found at ambient
pressure. 
It is plausible that the application of 
hydrostatic pressure to the polycrystalline material
has two effects. The first is on the 
structure of the triangular MnO planes as shown by neutron 
measurements \cite{kozlenko}.
At ambient pressures the Mn-O(3)-Mn and Mn-O(4)-Mn bond angles
and length differ slightly. This relieves the magnetic frustration
to an extent. The neutron diffraction measurements show that upon the 
application of pressure
the Mn-O(3)-Mn and Mn-O(4)-Mn bond angles and lengths approach each other.
These effects act to make the MnO planes more perfect realisations
of a triangular lattice, causing the exchange coupling along the triangular 
bonds to become more similar as pressure is increased. 
This increased frustration has been proposed as an explanation
for the reduction in magnetic moment \cite{kozlenko,janoschek}. 
The second effect is an increase in both $J'$ and $J$, which have
 an exponentially sensitive dependence
on bond distance, making the relative effect on $J'$ larger. 
INS measurements at ambient pressure show that $J'/J \sim 10^{-2}$
\cite{sato}, which is consistent with $T_{\mathrm{N}}/J
\sim 0.5$ ($J \approx 3$~meV \cite{park,sato}). Pressure will
therefore increase
$J'/J$ and hence $T_{\mathrm{N}}$.
It is clear that in order
for this dual effect to occur there exists a delicate balance
of competing interactions in this system.

This work was carried out at S$\mu$S, Paul Scherrer
Institut, Villigen CH and at the ISIS facility, Rutherford Appleton
Laboratory, UK.
We thank Hubertus Luetkens for technical assistance and EPSRC
(UK) for financial support. 
T.L.\ thanks the Royal Commission for the Exhibition of 1851
for support.


\begin{thebibliography}{99}
\bibitem{time}
Data were measured in 8192 bins, each of width 1.25~ns.
\bibitem{samples}
Samples were produced via a solid state reaction. 
Stoichiometric mixtures of Y$_{2}$O$_{3}$
and MnO$_{2}$ were ground together and calcined in air. For the 
polycrystalline material this process lasted 70 hours at
$\sim 1200^{\circ}$C with several 
intermediate grindings.
The single crystal material was pressed into a rod 
under 60~MPa hydrostatic pressure and crystals were grown in an
optical floating zone furnace (Crystal System Inc.) 
at a growth speed of ~5~mm/hour, with the seed and feed rods counter rotating
at 25 rpm. Purity was confirmed with X-ray diffraction.
\end{thebibliography}
\end{document}